\documentclass{aa}
\usepackage{graphics,graphicx}

\begin{document}

\title{Simultaneous single-pulse observations of radio pulsars}
\subtitle{II. Orthogonal polarization modes in PSR B1133+16}
\author{A. Karastergiou 
	\inst{1}
	\and
	M. Kramer
	\inst{2}
	\and
	S. Johnston
	\inst{3}
	\and
	A. G. Lyne
	\inst{2}
	\and
	N. D. R. Bhat
	\inst{4}
	\and
	Y. Gupta
	\inst{5}}

\institute{Max-Planck Institut f\"ur Radioastronomie, Auf dem H\"ugel
	69, 53121 Bonn, Germany \and Jodrell Bank Observatory,
	University of Manchester, Macclesfield, Chesire SK11 9DL, UK
	\and School of Physics, University of Sydney, NSW 2006,
	Australia \and Arecibo Observatory, HC3 Box 53995, Arecibo,
	Puerto Rico, PR 00612, USA \and NCRA, TIFR, Pune University Campus,
	Ganeshkhind, Pune 411007, India}

\abstract{In this paper, we present a study of orthogonal
polarization modes in the radio emission of PSR B1133+16, conducted
within the frame of simultaneous, multi-frequency, single-pulse
observations.  Simultaneously observing at two frequencies (1.41 GHz
and 4.85 GHz) provides the means to study the bandwidth of
polarization features such as the polarization position angle.  We
find two main results. First, that there is a high degree of
correlation between the polarization modes at the two frequencies.
Secondly, the modes occur more equally and the fractional linear
polarization decreases towards higher frequencies.  We discuss this
frequency evolution and propose propagation effects in the pulsar
magnetosphere as its origin.
 }

\maketitle

\section{Introduction}\label{intro}

Simultaneous multi-frequency observations of radio pulsars in full
polarization are proving to be a very powerful tool for investigating
the puzzling pulsar emission problem. In Karastergiou et
al. (\cite{khk01}, hereafter Paper I), we provided a detailed
introduction into the history of multi-frequency observations and the
ongoing efforts to construct an emission theory which will be both
self-consistent and consistent with the abundance of observational
information from pulsars. We also laid out the guidelines for
simultaneous polarimetric observations between different telescopes by
presenting our time-aligning and re-binning techniques.

Observations of the Vela pulsar by Radhakrishnan \& Cooke
(\cite{rc69}) showed the radiation to be highly linearly polarized and
to be emitted from near the magnetic pole. The observed position angle
(PA) reflects the instantaneous projection of the dipolar magnetic
field lines on the plane of the sky and thus changes with the rotation
of the star (the so-called rotating vector model, RVM).  Later,
Manchester et al. (\cite{mth75}) demonstrated that the percentage
polarization could vary between 0 and 100\% across pulse longitude,
that significant circular polarization was often present and, perhaps
most surprisingly, that significantly different PAs were seen at the
same pulse longitude.  These PA jumps were often very close to $90^o$
and thus the linear polarization occurred in `orthogonally polarized
modes' (OPM). The existence of OPM was a key discovery (Backer \&
Rankin 1980\nocite{br80}). By using simple arguments it can be shown
that the presence of OPM acting together in the emission process can
reduce the fraction of polarization observed (Stinebring et al. 1984,
McKinnon 1997, McKinnon \& Stinebring
1998\nocite{scr84}\nocite{mck97}\nocite{ms98}). Subsequently,
single pulse polarization studies by Gil \& Lyne (1995)\nocite{gl95}
demonstrated that OPM was responsible for the low percentage
polarization and very complicated PA variations in the integrated
profile of PSR B0329+54, in which two simple OPM modes existed.

Melrose \& Stoneham (\cite{ms77}) and Barnard \& Arons (\cite{ba86})
proposed a model for the OPM in pulsars by considering propagation
effects (refraction) which cause rays in the two natural modes of the
plasma to split. This has been further developed by McKinnon
(\cite{mck97}) who showed that a model involving birefringence of the
two modes can explain the observed pulse width and fractional
polarization changes with frequency.

In this paper, we shall focus on the polarization behaviour of
PSR~B1133+16 and use the results to draw some general conclusions
about the pulsar emission processes.

\begin{figure}[t]
\resizebox{\hsize}{!}{\includegraphics[angle=-90]{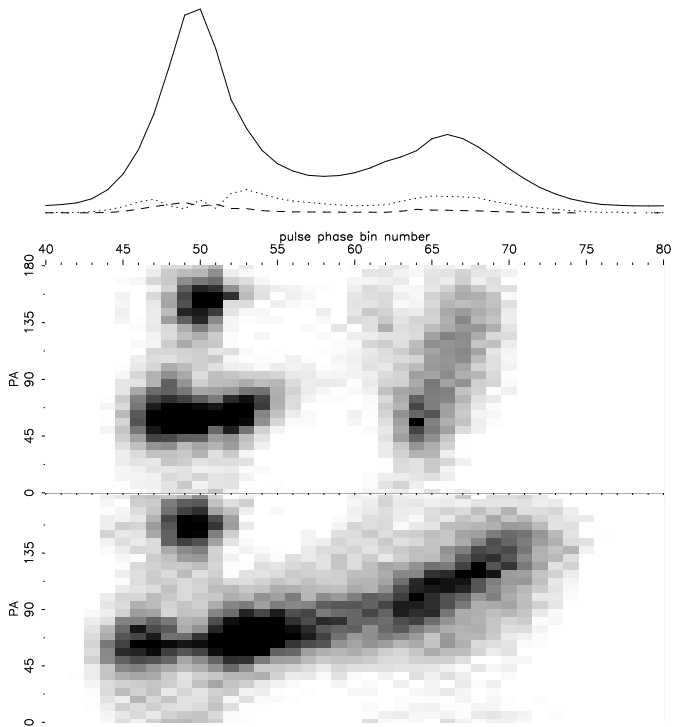}}
\caption[PAgray] {The total power profile of PSR B1133+16 is shown
together with the linear polarization at two frequencies; the
dotted line corresponds to 1.41 GHz and the dashed line to 4.85
GHz [TOP]. The grey-scale PA histograms follow, at 4.85 GHz [MIDDLE]
and 1.41 GHz [BOTTOM]. The darker areas in the grey-scale histograms
correspond to more frequent values of PA, as they occur on a
single-pulse basis. The PA values are not absolute.  }
\label{gray}
\end{figure}
\begin{figure}[t]
\resizebox{\hsize}{!}{\includegraphics[angle=-90]{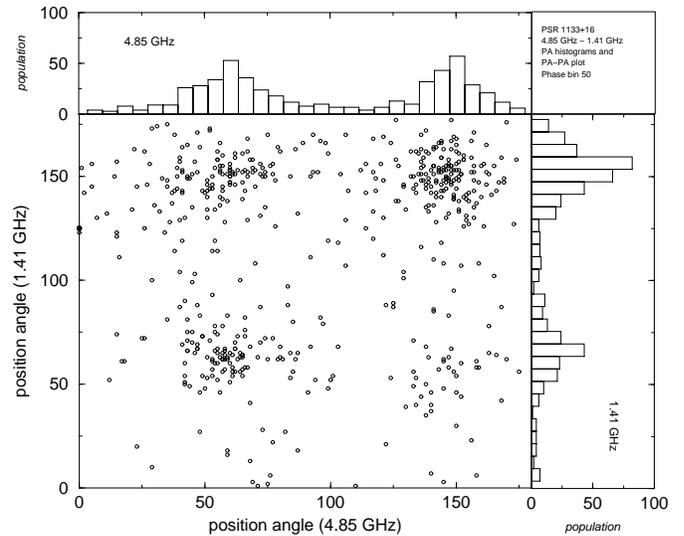}}
\caption[PA-PA plot] {PA-PA plot for phase bin 50. The distribution of
the PA values for each frequency is also plotted.}
\label{pahisto}
\end{figure}

\section{Observations}\label{Obs}

On 2000 January 4, we used the Effelsberg 100-m telescope at 4.85 GHz
and the Lovell 76-m telescope in Jodrell Bank at 1.41 GHz to observe
PSR B1133+16 (see von Hoensbroech et al. \cite{hkk98}, von Hoensbroech
\& Xilouris \cite{hx97} for details on the Effelsberg observing system
and Paper I and Gould \& Lyne \cite{gl98} for the Jodrell Bank
system). Both telescopes were simultaneously on source for $\approx
93$ minutes or 4778 pulse periods. For the purpose of this paper, we
also make use of Jodrell Bank data observed at 610 MHz (Gould \& Lyne
\cite{gl98}), as well as Effelsberg 2.69 GHz data (Paper I) from
different epochs.  The data from each telescope were calibrated and
converted into the EPN-format (Lorimer et al. \cite{ljs98}),
after being aligned and re-binned to a common reference frame and
temporal resolution (in this case 1.16 ms per bin), as described in
Paper I.
\section{Results}
\subsection{The behaviour of modes with frequency}
\begin{figure*}[t]
\resizebox{\hsize}{!}{\includegraphics[angle=-90,width=12cm]{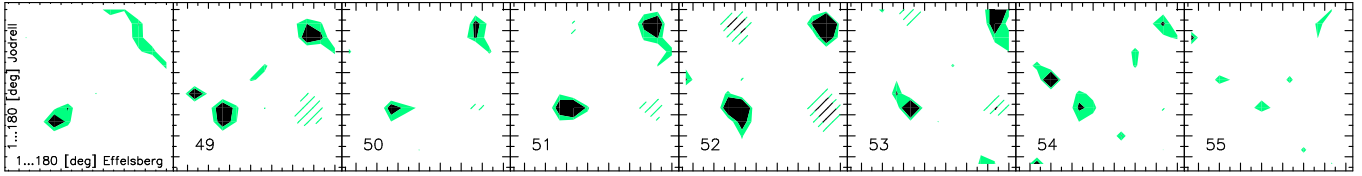}}
\caption[Modes and Predictions] {A sequence of phase bins from the
leading component of the pulse. Each frame consists of a PA-PA
histogram for the phase bin in the lower left corner, as identified in
Fig. 1, and shows the deviation of the observed PA pairs from the null
hypothesis, weighted by the calculated $\sigma$ (see explanation in
text).  }
\label{modes}
\end{figure*}

Fig. 1 shows the integrated pulse profile at 1.41 GHz along with the
linear polarization at both frequencies. In this paper we ignore the
circular polarization, which is generally low throughout the pulse.
Fig. 1 also shows the PA distribution as a function of phase in a
grey-scale representation at both 1.41 GHz and 4.85 GHz. It can be
seen that at the peak of the first component two orthogonal modes are
clearly present at both frequencies. In the second component one mode
dominates at 1.4 GHz but two modes can be seen at the higher
frequency. Integrated profiles like these have been used to question
the broad-band nature of OPM (e.g Stinebring et al. 1984); for the
first time, however, we can use the simultaneous nature of our data to
determine this issue.

At each of the two observed frequencies we compute the PA only if the
flux density of the linear polarization exceeds twice the noise.  We
do this for each phase bin of every pulse; an example of one such bin
near the peak of the first component is given in Fig \ref{pahisto}.
Under incoherent addition of the OPMs, the PA is only permitted to
take one of two discrete values, exactly $90^o$ apart. In this bin, it
is clear that the orthogonal PA modes are present at both frequencies,
although the spread in PA of each mode is somewhat larger than one
would expect from a delta-function convolved with instrumental noise
(Stinebring et al. 1984, McKinnon \& Stinebring
1998\nocite{scr84},\nocite{ms98}). Two clear results are apparent in
the figure. First, at 1.41 GHz the ratio of pulses in the two modes is
65:35 whereas at 4.85 GHz this ratio is 54:46. The modes are occurring
more equally at 4.85 GHz {\bf and} the fractional linear polarization
has also decreased. Secondly, all four possible combinations between
the two modes at the two frequencies occur, i.e.~four islands are
visible in Fig. \ref{pahisto}. Given the difference in the bimodal
distributions at the two frequencies, such a distribution of points in
the plot must arise, even in the case when there is no correlation
between the PAs of the two frequencies. Hence, it is not readily clear
from Fig.~2 whether the cases of agreement and disagreement of OPM,
represented by the four islands, suggest a high or low degree of
correlation between the frequencies. Obviously, the correlation is not
perfect since the individual distributions are different --- which is
an important result already -- but some degree of correlation may
still exist. We have to test whether the distribution of points in the
PA-PA plot can be attributed to some random combination of PA-values
at different frequencies, or whether the knowledge of the polarization
mode at one frequency allows one to predict the mode at the other to
some extent. We test the degree of correlation by comparing the
observed PA-PA distributions to the null hypothesis that the two
frequencies are completely uncorrelated.

We re-bin our PA distributions into histograms of 12 bins across the
$180^o$ range of PA values. By doing this we create a 2-dimensional
$12\times 12$ histogram $Obs(i,j)$, which represents the number of
occurrences of all possible PA-PA pairs as they are observed.  Then we
use Monte Carlo simulations (MCS) to randomly combine our two observed
PA distributions. We repeat this process a large number of times, thus
obtaining another 2-dimensional histogram $Null(i,j)$, similar to
$Obs(i,j)$, where the population in each square bin is the mean result
from all MCS iterations. At the same time, we can calculate the
standard deviation, $\sigma$, representing the statistical
fluctuations which we can expect in each square bin of the null
hypothesis case.  We now subtract the null hypothesis $Null(i,j)$ from
the observed 2-d histogram $Obs(i,j)$, and compare this difference
with the expected noise in each square bin, $\sigma$. This will
give us the significance in units of $\sigma$ that an observed
number of points $Obs(i,j)$ is not consistent with the null-hypothesis
of no correlation.

Such a null hypothesis test can only be applied reasonably to bins
where we observe {\em bimodal} PA distributions at {\em both}
frequencies, as other cases cannot be distinguished from a random
combination of PA values {\em per se}.  Hence, in Fig. \ref{modes} we
plot a 2-d PA-PA histogram only for the phase bins of the leading
pulse component where the individual PA distributions are bimodal.
The solid areas denote a difference between $Obs(i,j)$ and $Null(i,j)$
of more than $3 \sigma$ surrounded by a lighter coloured area marking
differences exceeding $2 \sigma$. Hatched areas denote differences of
less than $-3 \sigma$ surrounded again by lighter coloured area of $-2
\sigma$ significance.  We can clearly see that when one mode occurs at
one frequency, it also prefers to occur at the other frequency: the
same modes combine much more times than the null hypothesis of
no-correlation predicts (black areas on the x=y diagonal). 

In cases where the modes disagree, one mode does not necessarily
favour a value at the other frequency that is different by exactly
$90^o$. On the contrary, for most of the time the PA value at the
other frequency seems to be completely unrelated, resulting in a
distribution of points outside the diagonal islands that is mostly
consistent with the null hypothesis.  This is an interesting result
with possible implications on the widths of PA distributions at given
frequencies. Only in few cases is a lack of points observed when
compared to the null hypothesis (hatched areas), supporting an
increased degree of correlation between the frequencies. We note that
Fig. \ref{modes} does not change qualitatively by using grids coarser
than $12 \times 12$.

The results in this section therefore show that there is a high
degree of correlation between the two frequencies. The
same mode tends to dominate at both frequencies and
to this extent OPM is broadband in nature. At the same
time, the modes occur more equally and
the fractional polarization is lower at the higher frequency.
This will be discussed in more detail below.

\subsection{The $x$ and $s$ parameters}

According to Stinebring et al. (\cite{scr84}), the emission we receive
at any pulse phase is the instantaneous, incoherent superposition of
the two OPMs. The PA value is that of the strongest mode.  Two
important parameters can be constructed to investigate the strength
and behaviour of the modes.

The first parameter, $x$, represents the relative popularity of the PA
of one mode with respect to the other. By studying the individual PA
distributions for each frequency, it is possible to identify the OPMs
across the phase bins of the pulse. We can identify the PA of each
mode in each bin $i$ by finding the mean PA (or tracing
the PA swing when the mode seems absent) and identifying a PA range of
$\approx 30^o$ each side of the mean, which corresponds to the typical
width of the PA histograms, as shown in Fig. \ref{pahisto}.  $M_1$ is
the number of pulses where the PA corresponds to mode 1 and $M_{2}$
the same for mode 2.  For each of the frequencies we studied the
occurrence of $M_{1}$ and $M_{2}$ at every pulse phase bin.
\begin{figure}[t]
\resizebox{\hsize}{!}{\includegraphics{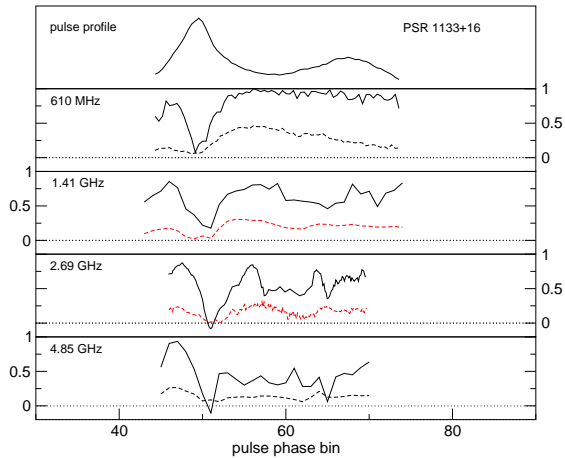}}
\caption[Ratios] {The parameter $x$ [solid line] and $|s|$ [dashed
line] across the phase bins of PSR B1133+16 for four different
frequencies.
There is a general trend for both $x$ and $s$ to become closer to zero
toward higher frequencies. 
}
\label{ratio}
\end{figure}

From the values of $M_1$ and $M_2$ we have calculated for each phase
bin, we construct 
$x=\frac{M_1-M_2}{M_1+M_2}$.
The value of $x$ depends on the relative values of $M_1$ and $M_2$. If
they are equal, $x$ will obviously be zero. On the other hand, if one
of them is larger than the other, $x$ will diverge from zero, 1 and $-1$
being the extreme cases, where only one of the modes is present. 

The second parameter is $s$, which represents the relative strengths
of the modes and varies from $-1$ to 1 in a similar fashion to
$x$. $s$ is defined as $s=\frac{S_1-S_2}{S_1+S_2}$, where each phase
bin is assumed to be the incoherent sum of two 100\% orthogonally
polarized modes with individual flux densities $S_1$ and $S_2$
(Stinebring et al. \cite{scr84}). In the case where the modes have a
negligible fraction of circular polarization (as is
the case here), then $|s|=L/I$, where
$L$ is the linear polarization and $I$ the total power.

We used the simultaneously observed data to study the behaviour of $x$
and $s$ with frequency. We noticed that, although they are both phase
dependent, they generally seem to have values closer to zero at 4.85
GHz than at 1.41 GHz (Fig. \ref{ratio}). We then used other
observations of PSR B1133+16 to see if this behaviour is consistent at
other frequencies. The Jodrell Bank 610 MHz data and the Effelsberg
2.69 GHz data, which are of comparable length to the simultaneous
data, clearly show this remarkable trend. In other words, the times
$M_2$ or $M_1$ determine the PA become more equal towards higher
frequencies, while at the same time mode 2 is becoming closer in
strength to mode 1 ($s$ closer to 0).

\section{Discussion}\label{disc}

In the case of PSR B1133+16 we have found the following trends of the
OPMs:

\begin{itemize}
\item OPMs have a spectral dependence: the polarized intensities of
the modes ($s$) and their frequency of occurrence ($x$) are becoming
comparable at higher frequencies;
\item we observe phase bins that simultaneously show a high value of
$x$ and a low value of $s$;
\item despite the spectral evolution, we find a strong
connection between the two frequencies: the same polarization modes
prefer to appear together;
\item different modes, however, do appear at the two frequencies,
implying a degree of de-correlation.
\end {itemize}
The first two trends are also supported by the 610 MHz and 2.69 GHz
data.

How well do these observational trends fit in with the model of
superposed orthogonal modes?
A high value of $x$ implies that a given mode dominates most of the
time. A low value of $s$ implies that the difference in the
average strengths of these two modes are rather small.
These two facts can only be reconciled by postulating that the
amplitudes of the two modes are highly correlated
(McKinnon \& Stinebring 1998). A physical mechanism which can
induce mode splitting with a high degree of amplitude correlation is
birefringence and this means
of mode creation in the relativistic pulsar
plasma was proposed by Barnard \& Arons (1986).

We also observe that, at least in this pulsar, $x$ decreases more
rapidly with frequency than does $s$. This clearly implies that the
correlation coefficient is decreasing with increasing frequency even
though the relative strengths of the modes change rather little.  One
observational consequence of this is that we would expect that the
distribution of linear polarization in the single pulses has a rather
low dispersion when the correlation coefficient is high, compared to
when the correlation coefficient is low.  This is indeed seen in the
data; the distribution of linear polarization is broader in the single
pulses observed at 4.85 GHz.

We therefore surmise that the relative strength of the modes is set by
the polarization of the input beam to the bi-refringent medium and
that this is largely constant with frequency. The frequency trend in
the observed value of $s$ arises from the refractive index of the
modes being frequency dependent - at some `critical' frequency and
above, the beams overlap, causing a decrease in $s$. In this pulsar,
$s$ changes rather little with frequency, as do the mode--separated
pulse widths between 1.41 GHz and 4.85 GHz
and presumably the critical frequency lies near 1.4 GHz. In most other
pulsars, however, the critical frequency seems to lie above 5 GHz
(Xilouris et al.\cite{xk96}).

The third and fourth conclusions we reach, together with the fact that
total power is well correlated in our data (as was the case for PSR
B0329+54 in Paper I), also point towards propagation effects.
They support the idea that OPMs, which could be created by the
splitting of the two natural orthogonal polarization modes in the
magnetosphere by a propagation effect like birefringence, endure
further propagation effects that change their relative strengths. One
explanation could be that the absorption (e.g. cyclotron absorption)
of one or other of the modes is frequency dependent with the higher
frequencies more affected than lower frequencies. A change in relative
strength could switch the dominant mode. The obvious manifestations of
this are the clearly seen OPM disagreements at the two frequencies.

A further observational test of propagation effects and the superposed
mode ideas comes from the circular polarization.  One expects that a
given mode is associated with a given sign of circular polarization
and that there is therefore a strong correlation between the dominant
mode and the sign of circular polarization in the single pulse
data. Given that we see a high correlation between mode occurrences at
different frequencies, will we also see a correlation between the
handedness of the circular polarization? Simultaneous,
multi-frequency, single--pulse observations are ideal as a test to
identify propagation effects at work on the radiation traveling
through the plasma in the pulsar magnetosphere.
\acknowledgements

We wish to thank Duncan Lorimer for his valuable suggestions on the
applied methodology. We also thank all the people working at the
participating telescopes who made these observations possible. Aris
Karastergiou also expresses his gratitude to the Deutsche Akademische
Austauschdienst for their financial support. We thank the referee,
Mark McKinnon, for suggestions on improving the paper.

\end{document}